\documentclass[12pt]{article}
\usepackage{epsfig,amsmath,graphics,bm}
\usepackage{graphics,graphicx}
\usepackage{tabularx}
\usepackage{amsfonts}
\usepackage{amsmath}
\usepackage{color}
\usepackage{float}
\usepackage{subcaption}
\usepackage{sidecap}

\def\ds{\displaystyle}

\newcommand{\beq}{\begin{equation}}
\newcommand{\eeq}{\end{equation}}
\newcommand{\lb}{\label}
\newcommand{\beqar}{\begin{eqnarray}}
\newcommand{\eeqar}{\end{eqnarray}}
\newcommand{\barr}{\begin{array}}
\newcommand{\earr}{\end{array}}

\newcommand{\derp}[2]{\ds{\frac {\partial #1}{\partial #2}}}

\def\scalp{\mbox{\boldmath$\, \cdot \, $}}

\def\XXint#1#2#3{{\setbox0=\hbox{$#1{#2#3}{\int}$}
     \vcenter{\hbox{$#2#3$}}\kern-.5\wd0}}

\def\c{{\circ}}

\def\bC{\mbox{\boldmath${\it C}$}}

\def\bD{\mbox{\boldmath${\it D}$}}

\def\bE{\mbox{\boldmath${\it E}$}}
\def\be{\mbox{\boldmath${\it e}$}}
\def\bF{\mbox{\boldmath${\it F}$}}

\def\bm{\mbox{\boldmath${\it m}$}}

\def\bn{\mbox{\boldmath${\it n}$}}
\def\b0{\mbox{\boldmath${\it 0}$}}

\def\bS{\mbox{\boldmath${\it S}$}}

\def\bx{\mbox{\boldmath${\it x}$}}

\def\Id{\mbox{\boldmath${\it I}$}}

\def\balpha{\mbox{\boldmath${\alpha}$}}

\def\Grad{{\rm Grad}}

\def\APL{{\em Appl. Phys. Lett.\ }}

\def\IJES{{ Int. J. Eng. Sci.\ }}

\def\JMPS{{\em J. Mech. Phys. Solids\ }}

\def\PRSL{{\em Proc. R. Soc. Lond.\ }}

\begin{document}

\title{Optimisation of hierarchical dielectric elastomer laminated composites\footnote{Published in International Journal of Non-Linear Mechanics,
Volume 106, November 2018, Pages 266-273, doi: 10.1016/j.ijnonlinmec.2018.06.005 }}

\author{
   Massimiliano Gei, Kudzai C.K. Mutasa
\\
\\
\small{\sl{School of Engineering, Cardiff University,}}\\
\small{\sl{The Parade, Cardiff CF24 3AA, Wales, U.K.}}\\
\small{\sl{Email: geim@cardiff.ac.uk}}}

\date{}
\maketitle

\begin{abstract}
This paper is concerned with the optimisation of the actuation response of electro-elastic, rank-two laminates obtained laminating a core rank-one composite with a soft phase which constitutes the shell. The analysis is performed for two classes of composites that are subjected to traction-free boundary-value problems. The results are compared with those computed in a previous study where the optimisation was carried out at small strains. The non-linear approach allows a better estimation of the geometric layout of the reference configuration to enhance the maximum stretch or shear strain at the operative applied voltage. The optimum layouts are in general characterised by a very low volume fraction of the shell while in the core the two components are almost equally distributed.
The amplification of the electric field in each phase of the laminate in the actuated state is also estimated to provide an indication of the effective local electro-mechanical response.
\end{abstract}


\section{Introduction}

The application of electro-elasticity theory of soft dielectrics \cite{dorf&ogde05acmc,mcmeeking} to the area of heterogenous materials has already shown that suitably designed microstructured composites may potentially exhibit remarkably improved actuation properties with respect to those displayed by the homogeneous materials employed so far, mainly acrylic elastomers and silicones \cite{gal_limor_mams07}--\cite{Bortot_jmps}. Some of these studies also investigated the shortcomings associated with the use of composites that are related to peculiar phenomena such as the triggering of micro-, macro-scopic and electro-mechanical instabilities, electric field amplification inside some constituents, etc. \cite{maxkatia2011,Rudykh,rudykh_apl,gei_lamin2013,rudykh_prsl_2014,spinelli_lopez_2015}.

A significant part of the research on heterogeneous soft dielectrics has been devoted to (hierarchical) layered composites. Working in the small-strain setting, Tian et al. \cite{Tian2012} have shown that two-phase dielectric rank-$n$ laminate layouts (with $n>1$) can be found to improve the electro-mechanical actuation strain of more than one order of magnitude at the same voltage (being the gain strongly related to the order $n$ of the microstructure). Or, in other words, that, in a rank-two laminate, the actuation enhancement obtained at an increasing contrast in the electro-mechanical properties of phases is more than proportional, almost linear in a graph where actuation strain and contrast are reported with a logarithmic scale.

In the more appropriate finite-strain, non-linear framework, laminated composites have been studied by some authors extending the methodology set out by deBotton \cite{gal2005lam}: deBotton et al. \cite{gal_limor_mams07} provided the first preliminary analysis of the
behaviour of rank-one laminates; Bertoldi and Gei \cite{maxkatia2011}, Rudykh and deBotton \cite{Rudykh}, Gei et al. \cite{gei_lamin2013}, Rudykh et al. \cite{rudykh_prsl_2014} and Spinelli and Lopez-Pamies \cite{spinelli_lopez_2015} investigated in detail the homogenisation and the micro- and macro-scopic stability of layered composites.
An evaluation of the benefit of the hierarchical structure of a rank-two composite in the electro-mechanical actuation has been carried out by
Rudykh et al. \cite{rudykh_apl} who have found a ten-fold improvement of the electro-mechanical coupling for a prototype laminate obtained by reinforcing with polyaniline an acrylic elastomer matrix.


This paper deals with the optimisation of the actuation response of rank-two dielectric laminates in the non-linear framework of finite electro-elasticity under both plane strain and three-dimensional unconstrained boundary conditions. The main goals are:

-- to extend the analysis of the optimisation of rank-two laminates presented in \cite{Tian2012}
to better assess the gain in large-strain actuation with respect to both the homogeneous and the rank-one response;

-- to establish the optimum configurations of the composite for different contrasts between the soft matrix and the stiff reinforcement;

-- to show that those configurations strongly depend on the maximum operational electric field chosen for the actuator;

-- to evaluate the amplification of the current electric field in some representative configurations to provide an indication of the effective local electro-mechanical response.


We are not concerned here with occurrence of any instability or other failure mode along the electro-mechanical actuation path, but we recognise that any departure from the homogeneous response may have a strong impact in the evaluation of the optimum layout.

Two material combinations are considered in the examples: one is that adopted by Tian et al. \cite{Tian2012}, characterised by a matrix whose electro-mechanical properties match those of a poly\-urethane, for which the authors analysed several configurations with equal contrast between shear moduli and electric permittivities; the other possesses a softer, silicone-like matrix, reinforced with a stiffer phase which may correspond to a poly(vinylidene fluoride) electrostrictive polymer \cite{Choi2016}.

\section{Background and homogenised response of laminated composites}
\label{section_mat}

\begin{figure}[t]
  \begin{center}
\includegraphics{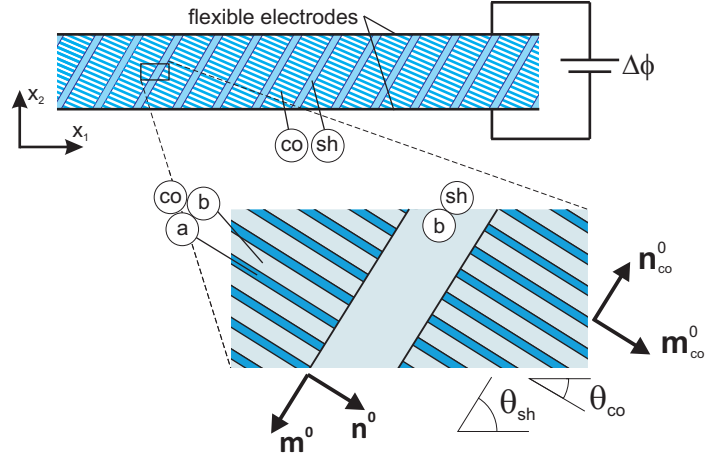}	
\caption{\footnotesize Geometry of the reference configuration $B^0$ of a two-phase, rank-two layered dielectric actuator subjected
to a voltage difference $\Delta \phi$ applied between electrodes. The close-up view represents the studied rank-two laminate highlighting the \emph{core} rank-one structure and the \emph{shell} composed of the soft matrix $b$. The initial thickness of the actuator is $h^0$. Axis $x_3$ is the out-of-plane direction, while angles $\theta_{\rm co}$ and $\theta_{\rm sh}$ are positive counter-clockwise.}
    \lb{geometry}
 \end{center}
\end{figure}

The general layout of the hierarchical composite actuator under investigation is represented in Fig. \ref{geometry}, where $h^0$ is the reference thickness.
The laminate is constructed layering two phases: $a$, the reinforcement, and $b$, the matrix,
usually softer than phase $a$ and with a lower electric permittivity,
in a manner consistent to the scheme called \lq tree a' in \cite{Tian2012}.
According to this scheme, the final layout is obtained layering a parent rank-one composite (the \emph{core}, abbreviated to \lq co', whose volume fraction will be denoted by $c^{\rm co}$) with a layer of soft material acting as a \emph{shell} (abbreviated to \lq sh', with volume fraction $c^{\rm sh}$ such that $c^{\rm sh}+c^{\rm co}=1$).
In \cite{Tian2012}, this layout has proven to give better performance than that of another type of arrangement, called \lq tree b'.
It is clear that while the materials to be mixed are two, namely, $a$ and $b$, the device is composed of three parts, i.e. $a$, $b^{\rm co}$ and $b^{\rm sh}$, with $b^{\rm sh}$ coinciding with the shell.
To complete the description of the notation related of volume fractions,
$c_a^{\rm co}$ and $c_b^{\rm co}$ denote, in turn, the volume fractions of $a$ and $b$ in the core,
whilst $c^a=c_a^{\rm co}$ and $c^b=c_b^{\rm co}+c^{\rm sh}$ are the total volume fractions of the two phases ($c^a+c^b=1$).
The grades with respect to
the axis $x_1$ of the interfaces between $a$ and $b$ in the core and of the shell correspond to $\theta_{\rm co}$ and $\theta_{\rm sh}$, respectively.
Separation of length-scales is assumed so that homogenisation at each rank can be invoked. The theory here recalled for the solution of the electro-elastic problem for hierarchical composites relies on the formulation that can be found in
\cite{ppcsiboni,Lopez2014,spinelli_lopez_2015,gal2005lam}, for which the effective properties can be obtained exactly up to a set of algebraic equations.

A notation which is standard in large-strain electro-elasticity is adopted\footnote{see \cite{gei_lamin2013}. Boldface letters $\bx^0$ and $\bx$ represent points of $B^0$ and of the current configuration $B$, respectively, while differential
operators with capital letters are evaluated with reference to $B^0$.}.
Symbols $\bD^0$ and $\bE^0$ denote, in general, nominal electric displacement and nominal electric fields, respectively, while $\bS$ and $\bF$ stand for the total first Piola-Kirchhoff stress tensor and the deformation gradient, respectively. As they were introduced, without any subscript, those four quantities are referred to the \emph{macroscopic} response of the composite whereas, when accompanied by a subscript,
they are referred to the indicated constituent so, e.g. $\bF_k=\Grad\, \bx_k\ (k=a, b^{\rm co}, b^{\rm sh})$ denotes the local deformation gradient that, together with the local nominal electric field $\bE^0_k$, is assumed to be constant in each part.
%
%
Both phases $a$ and $b$ obey an extended neo-Hookean strain-energy function formulated for incompressible materials, namely
\beq
    \lb{neohookeanE}
    W_k=\frac{\mu_k}{2}(\bF_k \scalp \bF_k-3)-\frac{\varepsilon_k}{2}(\bF_k^{-T} \bE_k^0\scalp \bF_k^{-T} \bE_k^0)\ \ \ \ (k=a, b^{\rm co}, b^{\rm sh}),
\eeq
where $\mu_k$, $\varepsilon_k$  represent the local shear modulus and electric permittivity, respectively. Clearly,
$W_{b^{\rm sh}}=W_{b^{\rm co}}=W_b$,
$\mu_{b^{\rm sh}}=\mu_{b^{\rm co}}=\mu_b$ and $\varepsilon_{b^{\rm sh}}=\varepsilon_{b^{\rm co}}=\varepsilon_b$.
In each part, the electro-elastic constitutive equations (see \cite{dorf&ogde05acmc}) provide the local response
\beq
        \lb{incomp_const_eq_phase}
        \bS_k=\derp{W_k}{\bF_k}-p_k\bF_k^{-T},\ \ \ \
        \bD^{0}_k=-\derp{W_k}{\bE_k^{0}} \ \ \ \ (k=a, b^{\rm co}, b^{\rm sh}),
\eeq
where $p_k$ is the arbitrary constant linked to the incompressibility constraint. Similarly to $\bF_k$ and $\bE^0_k$, $\bS_k$, $\bD^{0}_k$ and $p_k$ are constant in each part, as well.


The voltage $\Delta \phi$ applied between the
compliant electrodes will induce a macroscopic transverse electric field directed along $x_2$ whose nominal value $E^0$ corresponds to $\Delta \phi/h^0$, i.e. $\bE^0=\Delta \phi/h^0 \be_2$, where $\be_2$ is the unit vector directed along axis $x_2$.
The actuation response of the device can be achieved formally by solving two rank-one problems, that within the core and the macroscopic one involving core and shell. However, it is not necessary to solve in full the two tasks as we can exploit a remarkable result obtained by
Spinelli and Lopez-Pamies \cite{spinelli_lopez_2015} where an electro-elastic strain-energy function was directly formulated for a rank-one laminate, whose phases follow eq. (\ref{neohookeanE}), in terms of a set of invariants which takes into account the electro-mechanical anisotropic response of the material. This energy, here denoted by $W_{\rm co}$, has an expression reported in the Appendix.

Based on this observation, the macro-scale, rank-one problem involving core and shell (Fig. \ref{geometry}) can be solved by exploiting
the following jumps at the interface whose normal pointing toward the shell is $\bn^0$, namely,
$$
(\bF_{\rm co}-\bF_{\rm sh})\bm^0=\b0,\ \ \ \ (\bS_{\rm co}-\bS_{\rm sh}) \bn^0=\b0,
$$
\beq
\lb{jumpboundaryint}
(\bD^0_{\rm co}-\bD^0_{\rm sh})\scalp \bn^0=0,\ \ \ \ \bn^0 \times (\bE^0_{\rm co}-\bE^0_{\rm sh})=\b0,
\eeq
and the following constitutive equations, similar to \eqref{incomp_const_eq_phase},
$$
\bS_i=\derp{W_i}{\bF_i}-p_i\bF_i^{-T},\ \ \ \
        \bD^{0}_i=-\derp{W_i}{\bE_i^{0}} \ \ \ \ (i={\rm co}, {\rm sh}),
$$
where the notation is now obvious. It is only worth noting now that parameter $p_{\rm co}$ can be set as  $p_{\rm co}=p_a c^a+p_{b^{\rm co}}
c_b^{\rm co}$.

The relationships between local and macroscopic responses can be obtained by noting that, on the one hand, the underlying theory of composites at finite strain demands
    \beq
    \lb{eqMacroE}
    \bF=c^{\rm co}\bF_{\rm co}+c^{\rm sh}\bF_{\rm sh},\;\;\;\;
    \;\;\;\bE^{0}=c^{\rm co} \bE^{0}_{\rm co}+c^{\rm sh} \bE^{0}_{\rm sh};
    \eeq
on the other, interface compatibility \eqref{jumpboundaryint}$_1$ must be satisfied while \eqref{jumpboundaryint}$_4$ requires
   \beq
    \lb{E2alt}
    \bE^0_{\rm co}-\bE^0_{\rm sh}=\beta \bn^0.
    \eeq
As a consequence, admissible representations of $\bF_i$, $\bE^0_i$ ($i=\rm co,\, sh$) turn out to be
    \beq
    \lb{fafbav}
    \bF_{\rm co}=\bF\left(\Id+c^{\rm sh}\balpha\otimes\bn^0\right),\ \ \
    \bF_{\rm sh}=\bF\left(\Id-c^{\rm co}\balpha\otimes\bn^0\right)
    \eeq
and
\beq
    \lb{eabav}
    \bE^0_{\rm co}=\bE^0+c^{\rm sh}\beta \bn^0,\;\;\;\;\bE^0_{\rm sh}=\bE^0-c^{\rm co}\beta \bn^0,
\eeq
respectively.
In eq.  \eqref{fafbav}, $\balpha$ is a vector such that $\balpha \scalp \bn^0=0$.
In the problem addressed in the paper, for which $\bn^0 \scalp \be_3=0$, vector $\balpha$ is also orthogonal to $\be_3$,
then it is directed along the unit vector $\bm^0$ singling out the interface, i.e. $\balpha=\alpha \bm^0$.
Parameters $\alpha$ and $\beta$ are two fundamental unknowns of the problem under investigation that are associated with \eqref{jumpboundaryint}$_3$ and to the following scalar equation derived from \eqref{jumpboundaryint}$_2$
\beq
\lb{pippetto}
(\bS_{\rm co}-\bS_{\rm sh}) \bn^0\scalp \bm^0=0,
\eeq
respectively. The remaining unknowns are related to the macroscopic
response of the actuator whose kinematics is controlled by the deformation gradient that is assumed to be
\beq
\bF=\lambda \be_1\otimes\be_1+\xi/(\lambda\lambda_3) \be_1\otimes\be_2+1/(\lambda\lambda_3) \be_2\otimes\be_2+
\lambda_3 \be_3\otimes\be_3,
\eeq
where $\xi$ is the amount of shear related to the shear angle $\gamma$ in the current configuration ($\xi=\tan \gamma$) and, due to incompressibility, $\lambda_2=1/(\lambda \lambda_3)$. For plane strain conditions, $\lambda_3=1$.

The effective electro-elastic strain energy of the composite is the sum of the weighted energies of shell and core, namely
\beq
\lb{wav}
W=c^{\rm co}W_{\rm co}(\bF_{\rm co}(\alpha),\bE^0_{\rm co}(\beta),\bn^0_{\rm co})+c^{\rm sh}W_{b}(\bF_{\rm sh}(\alpha),\bE^0_{\rm sh}(\beta)),
\eeq
provided $\alpha$ and $\beta$ satisfy \eqref{jumpboundaryint}$_3$ and  \eqref{pippetto}. The macroscopic constitutive equations are \cite{ppcsiboni,Lopez2014}
  \beq
        \lb{incomp_const_eq_comp}
        \bS=\derp{W}{\bF}-p\bF^{-T},\ \ \ \
        \bD^{0}=-\derp{W}{\bE^{0}},
        \eeq
where the  lagrange multiplier $p$ can be taken as $p=p_{\rm sh} c^{\rm sh}+p_{\rm co} c^{\rm co}$.

From  eq. \eqref{wav}, it is clear that $W$ depends on the geometric and electro-mechanical properties of the microstructure, therefore, with a little abuse of notation\footnote{see the exact definition of $W$ in eq. \eqref{wav} and ensuing line.}, we can write
\beq
W(\lambda,\lambda_3,\xi,p,\alpha,\beta;K,E^0),
\eeq
where $K=\{c^{\rm co},c_b^{\rm co},\theta_{\rm sh},\theta_{\rm co},\mu_a,\mu_b,\varepsilon_a,\varepsilon_b\}$ and the list of six elements up to $\beta$ is the set of unknowns for the three-dimensional problem.
Once assigned geometry, properties of phases (i.e. the set $K$) and electrical input $E^0$, for the traction-free three-dimensional
boundary-value problems the six unknowns can be computed through the set of equations\footnote{throughout the paper, indices refer to the coordinate system introduced in Fig.  \ref{geometry}.}
\beq
S_{11}=S_{22}=S_{12}=S_{33}=0,\ \ \ (\bS_{\rm co}-\bS_{\rm sh}) \bn^0\scalp \bm^0=0, \ \ \ (\bD^0_{\rm co}-\bD^0_{\rm sh})\scalp \bn^0=0.
\eeq
Note that the shear stress $S_{32}$ is also vanishing, but this condition is here trivially satisfied due to the orientation of the laminae.
For plane-strain problems, $\lambda_3$ is not an unknown and equation $S_{33}=0$ is dropped from the solving system. Interestingly, from $p$,
multipliers $p_{\rm sh}$ and $p_{\rm co}$ in the two media can be attained from the definition of $p$ itself and the condition
$(\bS_{\rm co}-\bS_{\rm sh}) \bn^0\scalp \bn^0=0$ which provides an expression for their difference.

After having obtained all the electro-mechanical variables for the core, it is possible to compute the local quantities in the two phases $a$ and $b^{\rm co}$ following an approach similar to that described for the macro-scale core-shell rank-one problem. To this end, it is also possible to take advantage of the explicit expressions (14)--(16) obtained in \cite{gei_lamin2013} to find the parameters homologous to $\alpha$ and $\beta$ in the new problem and the pressure jump, respectively. The reader is referred to that reference for all details.

As shown in \cite{Lopez2014}, the total strain energy $W$ is formally such that it satisfies the min-max problem
\beq
W=\min_{\alpha} \max_{\beta} (c^{\rm co} W_{\rm co}+c^{\rm sh} W_b),
\label{minmaxW}
\eeq
where the sum in brackets corresponds to the right-end side of eq. \eqref{wav}.

Maximisation of the actuation quantity (i.e. either $\lambda$ or $|\xi|$) is performed at the maximum operational nominal electric field at which the actuator is supposed to work by employing the routine \texttt{FindMaximum} in the software package Mathematica (Wolfram Research, Inc., ver. 11.1). Due to the nonlinear nature of the problem, we have systematically and carefully checked the solutions of the various optimisation problems in the whole admissible domains of the involved variables. For layouts similar to those studied in \cite{Tian2012}, data provided in that reference are chosen as a starting point for the new search. In all analyses, we have checked that the parameters $\alpha$ and $\beta$ are such that the min-max condition \eqref{minmaxW} is fulfilled.


As recalled above, two types of matrix-reinforcement material combinations are adopted in the subsequent analyses:

 [{\bf composite no. 1}] that considered in \cite{Tian2012}, with a matrix $b$ with relatively \lq stiff' properties ($\mu_b=10$ MPa, $\varepsilon_b=10\, \varepsilon_0$), for which the authors studied contrasts $\mu^a/\mu^b=\varepsilon^a/\varepsilon^b=10,\ 100,\ 1000$ and $10000$;

 [{\bf composite no. 2}] that assembling a softer matrix ($\mu_b=0.1$ MPa, $\varepsilon_b=10\, \varepsilon_0$) with a reinforcing phase $a$ such that $\mu^a/\mu^b=10000$ and $\varepsilon^a/\varepsilon^b=100$.


\section{Some useful results relative to rank-one laminates}

The non-linear electro-elastic actuation behaviour of two-phase rank-one laminates has been well studied \cite{gal_limor_mams07,maxkatia2011,Rudykh,gei_lamin2013,spinelli_lopez_2015}. Here, we highlight a few new results for the traction-free boundary-value problems that complement the main analysis carried out for rank-two lamination layouts. In particular, for both types of composite introduced in the previous section, the two phases of the rank one displaying the highest actuation stretch share the same volume fraction (i.e. $c^a=c^b=0.5$).

For composite no. 1, in plane strain the lamination angle is $60.9^\c$ for all contrast ratios, while its amount is about $56^\c$ for the three-dimensional problem, again for all contrasts.
In Tian et al. \cite{Tian2012}, the best stretch for $\mu^a/\mu^b=\varepsilon^a/\varepsilon^b=100$ at an applied electric field of 100 MV/m was 1.0248 (in \cite{Tian2012},  the quantity that was actually computed was the actuation strain, here converted to stretch) with an angle of $61.2^\c$ (since for a homogeneous specimen, for which $c^b=1$, the stretch was 1.0221, the improvement amounted to $12.2\%$);
our finite-strain best stretch for the same laminate and applied nominal electric field is 1.0263. As homogeneous finite strain maximum stretch corresponds to 1.0234, the improvement is still in the order of $12.2\%$.

For the material with the \lq soft' matrix (composite no. 2), the plane strain optimum stretch at a nominal electric field of 20 MV/m corresponds to  1.1284
(
with an angle of the laminae of $63^\c$) while that for the three-dimensional actuator is equal to 1.1159
(same angle of $63^\c$); the homogeneous, plane strain counterpart amounts to 1.1154.

\section{Performance of traction-free rank-two laminates: composite no. 1}\label{inittt}

\subsection{Plane strain results with contrast parameter set to 100}

\begin{table}[t]
\footnotesize
	\centering
		\begin{tabular}{| c | c | c | c | c | c | c | c |}
			\hline
			Case& $\lambda_{\rm max}$ & $\xi$ & $c_b^{\rm co}$ & $\theta_{\rm co} [^{\circ}]$ & $c^{\rm co}$ & $\theta_{\rm sh} [^{\circ}]$\\
			\hline
			R2 T small str.& 1.1030 & not known & 0.531 & 60.3 & 0.964 & 14.6\\
			R2 T large str.& 1.1180 & 0.0724 & 0.531 & 60.3 & 0.964 & 14.6\\
			R2 opt (large str.)& 1.1241 & 0.1181 & 0.508 & 65.1 & 0.963 & 12.3  \\
\hline
			R2 opt -- 50 MV/m & 1.1203 & 0.0823 & 0.507 & 61.4 & 0.963 & 15.0\\
			\hline
	\end{tabular}
	\caption{Plane strain rank-two laminate configurations (composite no. 1) optimised for maximum stretch ($\lambda_{\rm max}$) for $\mu^a/\mu^b=\varepsilon^a/\varepsilon^b=100$ at $E^0=100$ MV/m (the first three rows). Variable $\xi$ is the corresponding amount of shear strain. \lq R2 T small str.' indicates the result presented by Tian et al. \cite{Tian2012} whilst \lq R2 T large str.' refers to the performance of the same configuration in the non-linear framework. \lq R2 opt' refers to the optimum layout at large strains. The fourth row refers to the performance at $E^0=100$ MV/m of the optimum layout computed at $E^0=50$  MV/m. Corresponding curves are presented on Fig. \ref{trer}.}
	\label{ranktwoaa}
\end{table}

\begin{figure}[t]
	\centering	
	\includegraphics[scale=0.7]{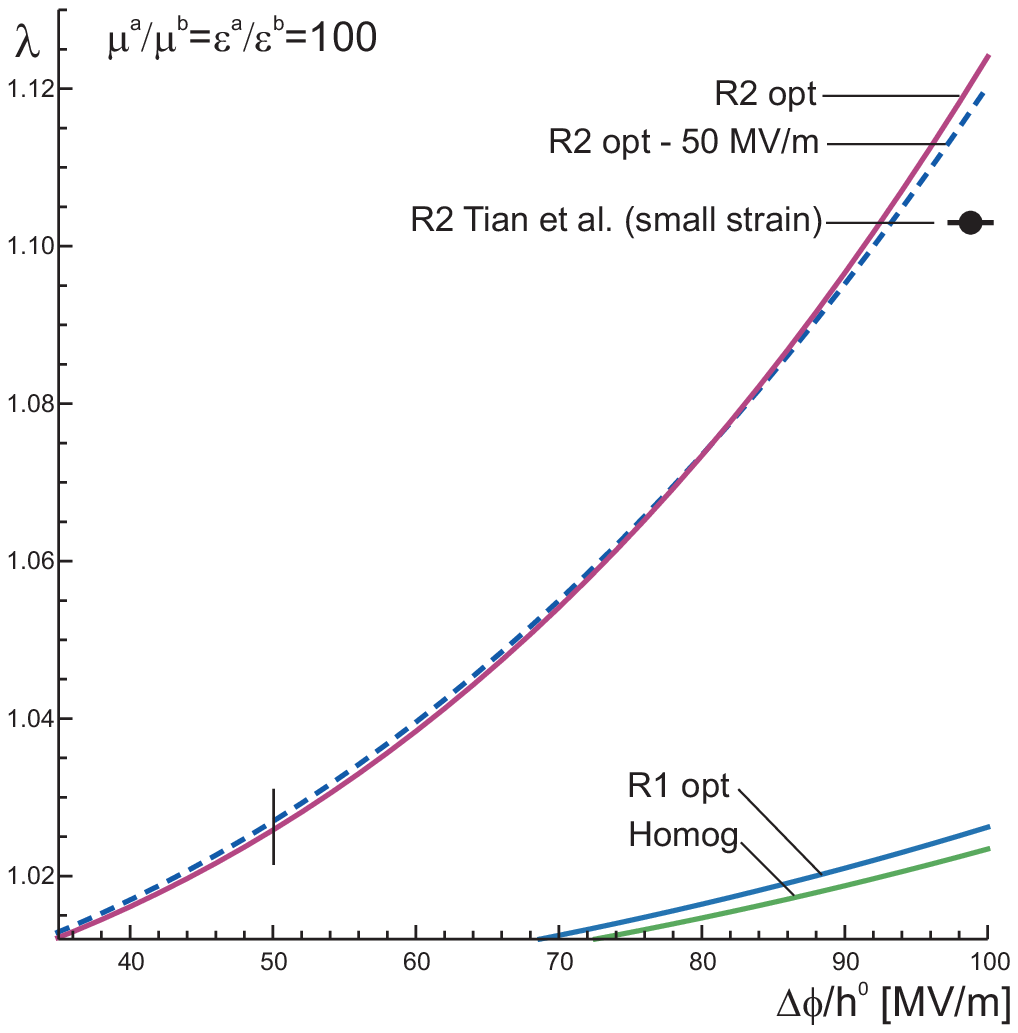}
	\caption{Longitudinal stretch (in plane strain) for optimum rank-two laminates (composite no. 1) at increasing applied nominal electric field for
$\mu^a/\mu^b=\varepsilon^a/\varepsilon^b=100$. Meaning of and configurations corresponding to \lq R2 opt', \lq R2 opt -- 50 MV/m' and \lq R2 T small str.' are displayed on Tab. \ref{ranktwoaa}. Curves relative to the homogeneous material ($c^b=1$) and optimised rank-one laminate are reported for comparison.}
	\label{trer}
\end{figure}

Our optimisation analysis for rank-two type actuators starts with composite no. 1 whose contrast parameter ($\mu^a/\mu^b=\varepsilon^a/\varepsilon^b$) is set to 100. The relevant results in terms of performance and geometry of the natural configuration are collected in Tab. \ref{ranktwoaa}, where the objective function of the optimisation is the maximisation of the longitudinal stretch ($\lambda_{\rm max}$) at a selected applied nominal electric field (100 MV/m for the first three rows, 50 MV/m for the fourth one).
In Fig. \ref{trer}, the longitudinal stretch functions in terms of the applied electric field are reported for the same configurations.
In Tab. \ref{ranktwoaa}, the first row refers to the results computed in \cite{Tian2012}, the second one reports the performance of the laminate layout obtained in \cite{Tian2012}, but taking into account the non-linear framework of the present approach.
\lq R2 opt' in the third row refers to the optimum layout at large strains.
This configuration shows an improvement in strain more than 4.7 times higher than the corresponding rank one, while it presents a 20\% improvement compared to the optimum configuration computed in \cite{Tian2012}. With respect to the latter, the lamination angle of the rank-one substructure (i.e. $\theta_{\rm co}$) has increased approximately by 5$^\c$ while that for the shell-core interface (i.e. $\theta_{\rm sh}$) has slightly decreased from 14.6$^\c$ to 12.3$^\c$.

The fourth row of Tab. \ref{ranktwoaa} describes the performance, at a nominal electric field of 100 MV/m, of a device whose geometry is optimised at $E^0=50$ MV/m (see the dashed line in Fig. \ref{trer}). Even though the difference in actuation of the two configurations is very small, this case shows that the optimisation of the geometric layout does depend on the operative electric field at which the device is designed to work.
The same phenomena would have been obtained for configurations optimised for any other nominal electric fields up to 100 MV/m. This confirms the presence of an optimum configuration that provides $\lambda_{\rm max}$ at a set applied nominal electric field.

As a final comment for this subsection, it is worth noticing that for the selected contrast parameters and in comparison to the findings of Tian et al. \cite{Tian2012}, the volume fraction of the rank-one core is almost unchanged while the volume fraction of the stiff phase in the core itself has slightly decreased from 0.531 to 0.508.


\subsection{Plane strain results: influence of increasing contrast}\label{sheer}


\begin{table}[t]
\footnotesize	\centering
		\begin{tabular}
			{| c | c | c | c | c | c | c | c |}
			\hline
			$\mu^a/\mu^b=\varepsilon^a/\varepsilon^b$& $\lambda_{\rm max} $& Gain w.r.t. hom.& $c^{\rm co}_b$ & $\theta_{\rm co} [^{\circ}]$ & $c^{\rm co}$ & $\theta_{\rm sh} [^{\circ}]$\\
			\hline
			10 & 1.0254 & 1.15 & 0.569 & 61.9 & 0.819 & 21.4\\
			100 & 1.1030 & 4.66 & 0.531 & 60.3 & 0.964 & 14.6\\
			1000 & 1.8272 & 37.43 & 0.584 & 63.1 & 0.992 & 27.5 \\
			10000 & 5.9849 & 225.56 & 0.690 & 62.8 & 0.997 & 42.3 \\
			\hline
	\end{tabular}
	\caption{Plane strain rank-two laminate configurations (composite no. 1) optimised for maximum stretch ($\lambda_{\rm max}$) in \emph{small strains} at $E^0=100$ MV/m as reported by Tian et al. \cite{Tian2012}. Different contrasts are considered. Values in the second column were originally presented as maximum strain and have been converted here in terms of stretch. The amount of shear is not available in \cite{Tian2012}. The gain in actuation strain with respect to the homogeneous actuator ($c^b=1$) is reported in the third column. Values of $\lambda_{\rm max}$ are sketched in Figs. \ref{trer} and \ref{trewddrd}.}
	\label{tablgtpp6}
\end{table}

\begin{table}
\footnotesize	\centering
		\begin{tabular}
			{| c | c | c | c | c | c | c | c | c |}
			\hline
			$\mu^a/\mu^b$& $\lambda_{\rm max} $ & Gain  &$\xi$ &Gain& $c^{\rm co}_b$ & $\theta_{\rm co} [^{\circ}]$ & $c^{\rm co}$ & $\theta_{\rm sh} [^{\circ}]$\\
			$=\varepsilon^a/\varepsilon^b$& & w.r.t. hom.& &  w.r.t. \cite{Tian2012}& & & & \\
            \hline
			10 & 1.0271 & 1.15&0.0013 &1.07& 0.563 & 62.1 & 0.831 & 20.1\\
			100 & 1.1241 & 5.29 &0.1181 &1.16 & 0.508 & 65.1 & 0.963 & 12.3\\
			1000 & 2.4789 & 63.10 & 0.2485 &1.26 & 0.498 & 70.0 & 0.990 & --20.0 \\
			10000 & 7.2227 & 265.47 &--7.9131 &1.17 & 0.518 & 88.0 & 0.988 & --51.0 \\
			\hline
	\end{tabular}
	\caption{Plane strain rank-two laminate configurations (composite no. 1) optimised for maximum stretch ($\lambda_{\rm max}$)  at $E^0=100$ MV/m in the non-linear framework.
Gains in actuation strain with respect to the homogeneous actuator ($c^b=1$) and the small strain assumption are respectively reported in the third and fifth column. Corresponding curves are presented on Fig. \ref{trewddrd}.}
	\label{tablgtpp}
\end{table}

\begin{figure}[t]
	\centering	
	\includegraphics[scale=0.7]{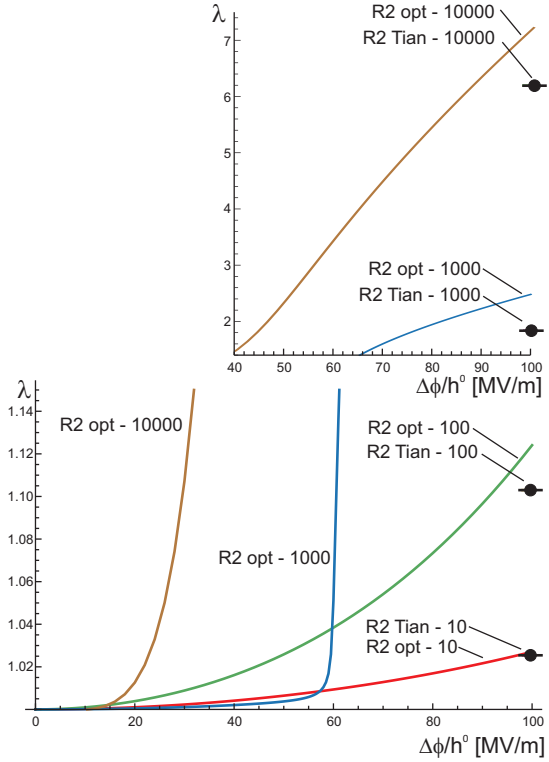}
	\caption{Longitudinal stretch (in plane strain) for optimum rank-two laminates (composite no. 1) at increasing applied nominal electric field for different contrasts. The markers at $E^0=100$ MV/m represent values obtained by Tian et al. \cite{Tian2012} in small strains (see Tab. \ref{tablgtpp6}). The sub-image displays curves for contrasts 1000 and 10000 at high electric field. Corresponding configurations are presented on Tab. \ref{tablgtpp}.}
	\label{trewddrd}
\end{figure}

Table \ref{tablgtpp6} shows the results obtained in \cite{Tian2012} for contrasts up to 10000 at small strains along with Tab. \ref{tablgtpp} which presents data obtained by optimisation of the maximum stretch with the non-linear approach. The second row of the latter corresponds to that labelled \lq R2 opt (large str.)' in Tab. \ref{ranktwoaa}.

At the lowest contrast of 10, the beneficial effect of the hierarchical composite was found to be less pronounced (longitudinal stretch equal to 1.0271 from 1.0263 of the rank one) such that the improvement in the rank-two laminate from the rank one is very small.
A substantial increase in $\lambda_{\rm max}$ from the homogeneous value has been obtained at contrasts of 100 and higher and this reflects a remarkable improvement in maximum longitudinal stretch attainable at large strains.
The trend of the improvement with contrast follows a similar rule to that highlighted in \cite{Tian2012}, however the performance obtained with the non-linear model is, on average, 18\% higher than that predicted at small strains.

The computations carried out in the present paper confirm that the optimum composite possesses a very limited volume fraction of the soft shell material (whose volume fraction is equal to $1-c^{\rm co}$), whilst, opposite to the finding presented in \cite{Tian2012}, the volume fraction of the soft phase in the inner rank-one phase remains approximately about 0.5, independently of the value of the contrast parameter.
We also note that in the non-linear setting the improvement in actuation stretch with respect to the homogeneous response is more correctly estimated and is always higher than that computed with a small-strain formulation (up to 70\% higher for the contrast 1000).

A sensitivity analysis on how changes in lamination angles $\theta_{\rm co}$ and $\theta_{\rm sh}$ impact on the actuation stretch has been carried out revealing that the former angle plays the most important role. This may explain why, by comparing angles in Tabs. \ref{tablgtpp6}
and \ref{tablgtpp}, the grades of the interfaces within the core are similar at the same contrast, while angles $\theta_{\rm sh}$ are very different at high stiffness and permittivity ratios (1000 and 10000). The increase of $\theta_{\rm co}$ with contrast in Tab. \ref{tablgtpp} can be explained bearing in mind that, at large strains, one of the components of the deformation for all phases is the rotation, then the initial geometry strongly depends on the final configuration of the actuated state.

Fig. \ref{trewddrd} presents graphically the behaviour of the laminates listed in Tabs. \ref{tablgtpp6} and \ref{tablgtpp} as a function of the applied nominal electric field. As the stretch $\lambda$  scales across two orders of magnitude, the picture is composed of two plots.
The amount of shear strain accompanying the longitudinal stretch is however considerably high at a contrast of 10000, which may imply that it is almost impossible to reach the predicted maximum stretch without reaching some material failures  associated with strong through-thickness strain gradients, i.e. interface debonding.
Therefore, in all tables of this paper, we may see the figures of the optimisation for $\mu^a/\mu^b=\varepsilon^a/\varepsilon^b=10000$ as purely indicative of the outcome of the algorithm, but with a limited practical applicability.

\begin{figure}[t]
	\centering	
	\includegraphics[scale=0.7]{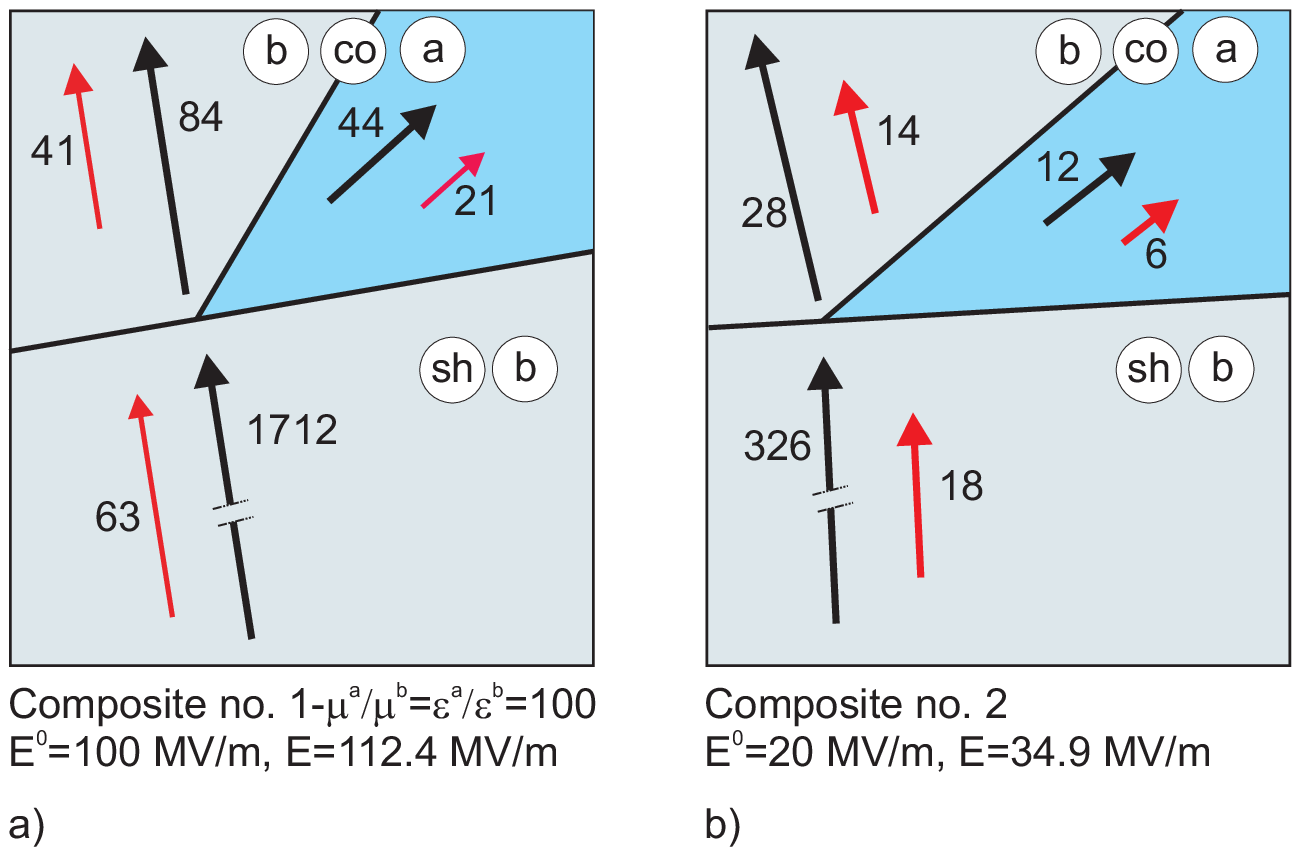}
	\caption{Schematics showing the current electric fields (in black) and their weighted contributions (in red) to the macroscopic electric field ($\bE$) in the three phases of the composite (values in MV/m):
a) composite no. 1 with $\mu^a/\mu^b=\varepsilon^a/\varepsilon^b=100$ at $E^0=100$ MV/m ($E= 112.4$ MV/m); b) composite no. 2 ($E= 34.9$ MV/m in this case). Electric fields in phase $b^{\rm sh}$ are not to scale.}
	\label{current_electric}
\end{figure}

As a part of the solution, we can track the evolution of the current configuration of the composite along the electro-mechanical actuation. Here, we report some results for contrasts 100 and 1000 in Tab. \ref{tablgtpp}. At the maximum applied electric field, for the former, $\theta_{\rm co}=59.2^\c$ and $\theta_{\rm sh}=9.6^\c$; for the latter case, the two angles reach respectively the values of 24.1$^\c$ and $-3.4^\c$.
At the same final actuated configurations, the current electric field in the different phases can be computed exploiting expression (see \cite{dorf&ogde05acmc,maxkatia2011})
$
\bE_k=\bF_k^{-T}\bE_k^0 \ \ (k=a, b^{\rm co}, b^{\rm sh}).
$
This will help to have a complete overview of the current status of each component of the composite.
For the contrast 100, the intensity of the current electric field in the shell (composed of the matrix, phase $b$) is 1712 MV/m, that is about 15 times larger than that of the macroscopic average electric field in the actuator ($E$) that corresponds to 112.4 MV/m. In Fig. \ref{current_electric}a, the electric fields in the three constituent phases are reported, together with the weighted counterparts whose sum corresponds to the current overall electric field $\bE$ that, as $\bE^0$, is directed along the thickness of the actuator.

Optimisation of laminates in terms of $\xi_{\rm max}$ can also be performed providing crucial insight into the effective influence of microstructural interactions (data collected in Tab. \ref{tablgtpdp} where the companion longitudinal stretch $\lambda$ is also reported). The procedure has been carried out with the objective to maximise
$|\xi_{\rm max}|$ at $E^0=100$ MV/m, so that, in the table, the amount of shear can be either positive or negative.

\begin{figure}[t]
	\centering	
	\includegraphics{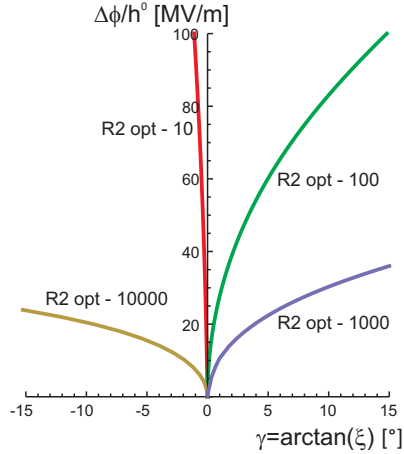}
	\caption{Shear angle for optimum rank-two laminate layouts in plane strain (composite no. 1) at increasing applied nominal electric field for different contrasts. Corresponding configurations are presented on Tab. \ref{tablgtpdp}.}
	\label{trewddrdi}
\end{figure}

Fig. \ref{trewddrdi} displays the applied nominal electric field influence on shear angle $\gamma=\arctan(\xi)$ (in the range $-15^{\circ}\leq\gamma\leq15^{\circ}$) and we observe that the behaviours of the different curves are strongly scattered.
For instance, at contrasts of 100 and 1000 a positive overall maximum shear strain takes place, implying that the material is shearing in a clockwise direction, while the behaviour is opposite for the other two reported cases. We also note that for the two higher contrasts the laminate diverges to high shearing angles at a relatively low electric excitation.

For the limited number of cases analysed, the lack of a general trend in the shear strain optimisation is also clear from the lamination angles obtained at different stiffness and permittivity ratios. Interestingly, at the lowest contrast, $\theta_{\rm co}=\theta_{\rm sh}$, so that the most efficient arrangement of phases corresponds to a rank-one layout.
For $\mu^a/\mu^b=\varepsilon^a/\varepsilon^b=100, 1000$, the layers of the sub rank-one structure of the core are at a right angle with respect to $x_1$.

\begin{table}
\footnotesize	\centering
		\begin{tabular}
			{| c | c | c | c | c | c | c | c |}
			\hline
			$\mu^a/\mu^b$ & $\xi_{\rm max} $& $\lambda $& Gain in $\lambda$   & $c^{\rm co}_b$ & $\theta_{\rm co} [^{\circ}]$ & $c^{\rm co}$ & $\theta_{\rm sh} [^{\circ}]$\\
            $=\varepsilon^a/\varepsilon^b$ &  & & w.r.t. homog. & & &  & \\
			\hline
			10 & --0.0201 &1.0174 & 0.74 & 0.499 & 30.8 & 0.998 & 30.8\\
			100 & 0.2642 & 1.0617 & 2.63 & 0.505 & 90.0 & 0.967 & 31.1\\
			1000  & 1.7290& 2.2419 & 46.62 & 0.494 & 92.0 & 0.994 & 27.0\\
			10000 & --20.904 &7.0147 & 256.60& 0.529 & 58.8 & 0.980 & 90.0 \\
			\hline
	\end{tabular}
	\caption{Plane strain rank-two laminate configurations (composite no. 1) optimised for maximum amount shear deformation ($\xi_{\rm max}$)  at $E^0=100$ MV/m. Different contrasts are considered. The gain in actuation strain with respect to the homogeneous actuator ($c^b=1$) is reported in the fourth column. Corresponding curves are presented on Fig. \ref{trewddrdi}.}
	\label{tablgtpdp}
\end{table}

\subsection{Performance of rank-two laminates in three dimensions with vanishing tractions}
\label{sshhea}

We will briefly analyse here the performance of rank-two actuators in a three-dimensional environment where nominal tractions are all vanishing at the boundaries. As laminae at different hierarchical order are still aligned parallel to axis $x_3$,
the new optimum configurations, whose data are collected in Tab. \ref{tabdlq}, are very similar to those obtained for plane-strain actuation (Tab. \ref{tablgtpp}), with the out-of-plane stretch $\lambda_3$ being very close to 1. Only for the contrast 10000, the two angles $\theta_{\rm sh}$ are very different. However, as $\theta_{\rm co}$ is almost at a right angle for both, the two different inclinations of the shell are almost symmetric with respect to axis $x_1$ so that the two configurations provide almost the same longitudinal stretch. The stretches $\lambda_{\rm max}$ in the two tables are very similar, with those computed for three-dimensional boundary-value problems being slightly higher at high stiffness and permittivity ratios.

\begin{figure}[thb]
	\centering
	\begin{subfigure}{0.45\textwidth}
		\centering
		\includegraphics[width=1\linewidth]{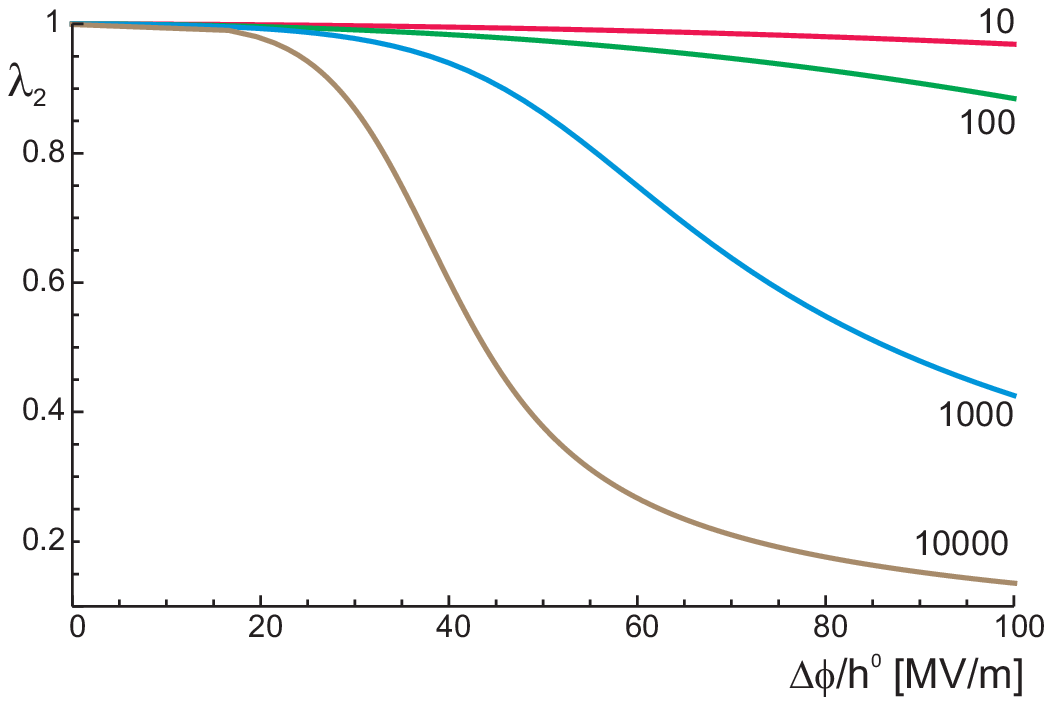}
		\caption{}
		\label{lambdathreedeetwosthearq}
	\end{subfigure}%
\hspace{5 mm}
	\begin{subfigure}{0.45\textwidth}
		\centering
		\includegraphics[width=1\linewidth]{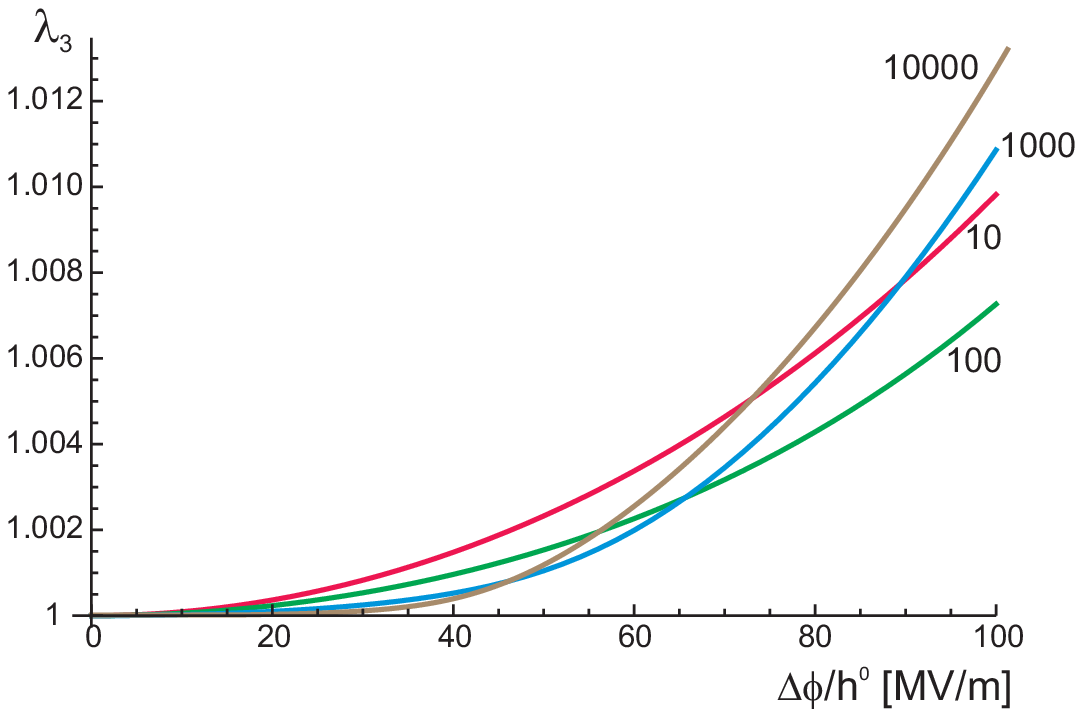}
		\caption{}
		\label{lambdathreedeethreetshearq}
	\end{subfigure}
	\caption{Influence of increasing applied electric field on (a) transverse stretch ($\lambda_2$) and (b) out-of-plane stretch ($\lambda_3$) for rank-two laminates (composite no. 1) relating to a three dimensional boundary-value problem. Values of contrasts are reported close to each curve. Configurations are presented on Tab. \ref{tabdlq}.}
	\label{lambda2d3dR2eetdrq}
\end{figure}

As the longitudinal stretch-electric field path $\lambda-E^0$ in the three-dimensional framework is very similar to that for plane strain, only transverse stretch and out-of-plane stretch actuation paths are reported in Fig. \ref{lambda2d3dR2eetdrq}, in parts a) and b), respectively. In Fig. \ref{lambda2d3dR2eetdrq}b, the stretch $\lambda_3$ corresponding to a contrast of 10 is unexpectedly higher than those of the other samples up to approximately $E^0=70$ MV/m, remaining higher than that of the case 100 for the whole actuation range.

\begin{table}[t!hb]
\footnotesize
	\centering
		\begin{tabular}
			{| c | c | c | c | c | c | c | c | c | c |}
			\hline
			$\mu^a/\mu^b$& $\lambda_{\rm max}$  & $\lambda_2$ & $\lambda_3$ & $\xi$& $c^{\rm co}_b$ & $\theta_{\rm co} [^{\circ}]$ & $c^{\rm co}$ & $\theta_{\rm sh} [^{\circ}]$\\
			$=\varepsilon^a/\varepsilon^b$ &  &  &  & &  &  & & \\
\hline
			10  & 1.0223 & 0.9687 & 1.0098 & 0.0004 & 0.562 & 55.4 & 0.824 & 5.5\\
			100  & 1.1221 & 0.8848 & 1.0073 & 0.1142 & 0.508 & 64.4 & 0.964 & 11.0\\
			1000 & 2.4890 & 0.3978 & 1.0099 & 0.2369 & 0.510 & 70.0 & 0.989 & --20.0\\
			10000 & 7.3097 & 0.1351 & 1.0131 & 6.2339 & 0.496 & 90.0 & 0.990 & 47.0\\
			\hline
	\end{tabular}
	\caption{Three-dimensional rank-two laminate configurations optimised for maximum stretch ($\lambda_{\rm max}$)  at $E^0=100$ MV/m in the non-linear framework with
transverse ($\lambda_2$) and  out-of-plane ($\lambda_3$) stretches determined. Corresponding curves are displayed on Fig. \ref{lambda2d3dR2eetdrq}.}
	\label{tabdlq}
\end{table}

Similarly to the actuation stretch $\lambda$, optimisation with respect to $|\xi_{\rm max}|$ has been also carried out for a three-dimensional boundary value problem. Again, due to the high out-of-plane stiffness of the actuators, $\lambda_3$ is only slightly higher than 1 for all contrasts and the resulting optimum layouts are very similar to those obtained in the corresponding plane strain problems.

\section{Performance of traction-free rank-two laminates: composite no. 2}

\begin{table}[b!]
\footnotesize	
\centering
		\begin{tabular}
			{| c | c | c | c | c | c| c | c | c | c |}
		\hline
		Case & $\lambda_{\rm max}$ & $\lambda_2$ & $\lambda_3$ & $\xi$ & $c^{\rm co}_b $ & $\theta_{\rm co} [^{\circ}]$ & $c^{\rm co}$ & $\theta_{\rm sh} [^{\circ}]$\\
		\hline
		2D (opt $E^0=20$ MV/m)& 1.7588 &0.5686 & 1 & --1.329 &0.520 & 88.0 & 0.944 & 14.0\\
		3D (opt $E^0=20$ MV/m) & 1.8483 & 0.5166 &1.0474 & --0.0292 &0.517 & 59.0 & 0.957 & --32.0\\
			\hline
      2D (opt $E^0=10$ MV/m) & 1.6503 & 0.6060 &1 & 0.1195 &0.508 & 65.1 & 0.964 & 12.0\\
      3D (opt $E^0=10$ MV/m) & 1.6924 & 0.5612 &1.0528 & 0.0704 &0.509 & 65.0 & 0.963 & 11.9\\
			\hline
	\end{tabular}
	\caption{Rank-two laminate configurations and corresponding maximum longitudinal stretch (composite no. 2) for an actuation nominal electric field of $E^0=20$ MV/m. First (last) two rows: the composite is optimised for $E^0=20$ MV/m ($E^0=10$ MV/m). Corresponding curves are presented on Figs. \ref{rankoowttqitq1} and \ref{trewddrr4rcs}.}
	\label{tablpdpd}
\end{table}

\begin{figure}[t!]
	\centering
	\includegraphics{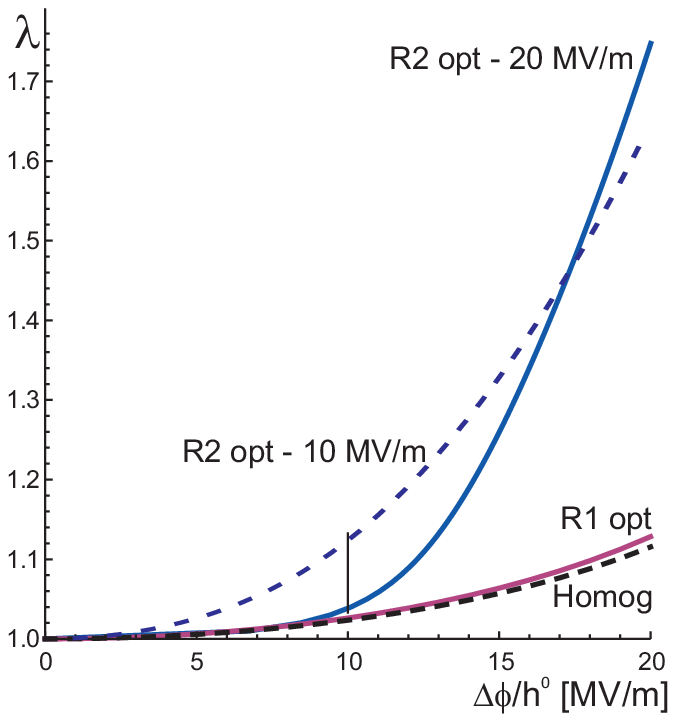}
	\caption{Longitudinal stretch (in plane strain) for optimum rank-two laminates (composite no. 2) at increasing applied nominal electric field. Both optimal configurations for 10 MV/m and 20 MV/m are reported. Curves relative to the homogeneous material ($c^b=1$) and optimised rank-one laminate are displayed for comparison. Corresponding configurations are presented on Tab. \ref{tablpdpd}.}
	\label{rankoowttqitq1}
\end{figure}

The optimum rank-two laminate layouts for the composite no. 2 are collected in Tab. \ref{tablpdpd} for both plane strain and three-dimensional traction-free problems (optimisation of $\lambda_{\rm max}$). Guided by the study of rank-one laminates by Gei et al. \cite{gei_lamin2013}, we have selected, at an initial stage, 20 MV/m as the maximum applied nominal electric field. For this level of electric excitation, it is confirmed that the most effective microstructure arrangement corresponds to the one where the soft shell has a relatively low volume fraction (equal to 0.056 in plane strain and to 0.043 in 3D), while the two phases are almost equi-distributed in the sub rank-one
core. In plane strain, the optimum $\theta_{\rm co}$ is very close to the right angle and the actuation is accompanied by a large shear strain. The plots in Fig. \ref{rankoowttqitq1} show that the obtained rank two displays an actuation strain that is 6.6 times higher than that of the homogeneous actuator without reinforcement, while the associated optimum rank one only slightly outperforms the homogeneous response.
Also for the composite no. 2, it is possible to track the change in angles $\theta_{\rm co}$ and $\theta_{\rm sh}$ along the electro-mechanical actuation path. For the plane strain case, the evolution of the angles is quite smooth; $\theta_{\rm co}$ will reduce from 88$^\c$ to 40$^\c$ at 20 MV/m, while $\theta_{\rm sh}$ will decrease from 14$^\c$ to about 4$^\c$.
The set of current electric fields in the composite at $E^0=20$ MV/m is represented in Fig. \ref{current_electric}b. That in the soft shell reaches 325.6 MV/m, with an amplification factor of approximately 9.3 with respect to the intensity of the overall average electric field ($E$) experienced by the composite, whose value is 34.9 MV/m.

Fig. \ref{rankoowttqitq1} also shows the performance of an actuator optimised at a working nominal electric field of 10 MV/m whose layout information are listed in Tab. \ref{tablpdpd}.  At this electric field, the device under investigation clearly outclasses that optimised for 20 MV/m, being the longitudinal strain of the former more than three times larger than that of the latter. This feature obviously reverses at 20 MV/m.

These computations show the importance of the chosen working electric field for the optimisation of the composite layout.
For 20 MV/m, in the three-dimensional case, the two initial layering angles are quite different than those computed in plane strain. The reason is that, differently from composite no. 1, the out-of-plane strain is now not negligibly small, then the two studied boundary-value problems lead to different layouts. On the contrary, for 10 MV/m the two configurations are very similar as at this nominal electric field the stretch $\lambda_3$
is very close to 1 in 3D case.

\begin{figure}[t]
	\centering
	\begin{subfigure}{0.5\textwidth}
		\centering
		\includegraphics[width=1\linewidth]{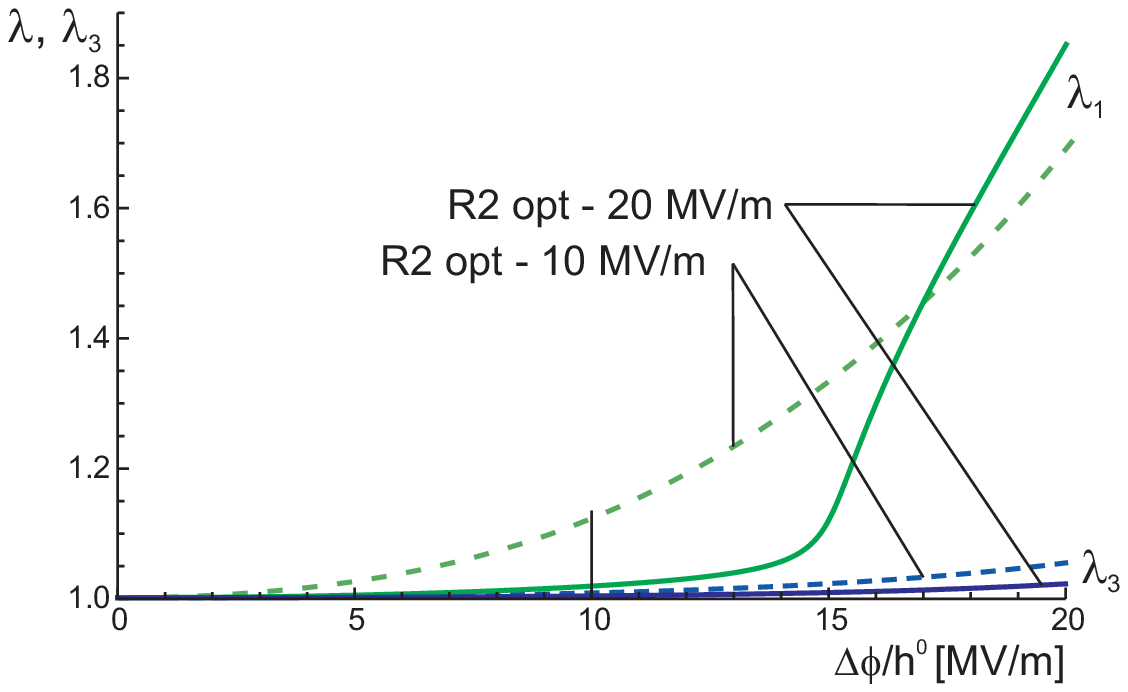}
		\caption{}
		\label{asascs}
	\end{subfigure}%
\hspace{5 mm}
	\begin{subfigure}{0.44\textwidth}
		\centering
		\includegraphics[width=1\linewidth]{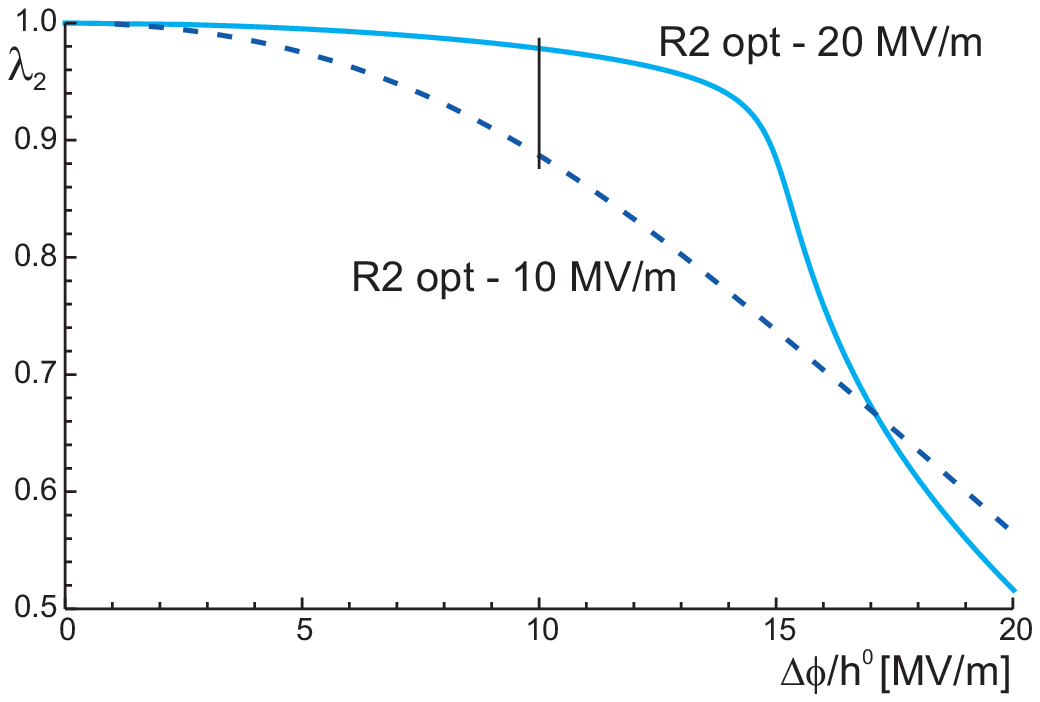}
		\caption{}
		\label{asacs}
	\end{subfigure}
	\caption{Influence of increasing applied electric field on (a) longitudinal ($\lambda$) and out-of-plane ($\lambda_3$) stretches  and (b) transverse stretch ($\lambda_2$) for a traction-free rank-two composite with \lq soft' matrix (composite no. 2) in the three-dimensional boundary-value problem. Both optimal configurations for 10 MV/m and 20 MV/m, whose parameters are presented on Tab. \ref{tablpdpd}, are reported.}
	\label{trewddrr4rcs}
\end{figure}

Fig. \ref{trewddrr4rcs} reports the three stretches along actuation for the two three-dimensional problems. It can be noticed that the response of the actuator, for that optimised at 20 MV/m, is very poor up to the value $E^0\approx 14.5$ MV/m, beyond which the system becomes suddenly very compliant. The same trend is followed by the angles $\theta_{\rm co}$ and $\theta_{\rm sh}$: they remain almost constant up to $E^0\approx 14.5$ MV/m and then change quickly to reach respectively the values of 24.9$^\c$ and $-9.8^\c$.
For the layout optimised at $E^0=$ 10 MV/m, the increase in actuation is smoother and more similar to a parabolic function.

\section{Conclusions}

The problem of actuation optimisation of hierarchical rank-two dielectric elastomer composites is tackled in the non-linear electro-elasticity framework for both plane strain and three-dimensional traction free boundary-value problems.
The results complement and extend those reported for small-strain computations by Tian et al. \cite{Tian2012}.

Two types of composites are considered: one is that adopted by Tian et al. \cite{Tian2012}, characterised by a matrix whose electro-mechanical properties match those of a polyurethane, for which the authors analysed configurations with equal contrast between shear moduli and electric permittivities; the other possesses a softer, silicone-like matrix, reinforced with a phase which may correspond to a poly(vinylidene fluoride) electrostrictive polymer \cite{Choi2016}.

The following conclusions can be drawn from this study.

The non-linear framework allows a better estimation of the optimum layouts of the composite that deform typically at large strains.
For the former type of composite, the trend of the actuation enhancement with contrast follows a similar rule to that highlighted in \cite{Tian2012}, however the performance obtained with the non-linear model is, on average, 18\% higher than that predicted at small strains. As far as the volume fractions are concerned, that of the soft phase in the inner sub rank-one core remains approximately about 0.5, independently of the value of the contrast parameter.
For all the analysed cases, the optimum composite possesses a very limited volume fraction of the soft shell material (in the order of a few percents) where the electric field reaches an intensity that is one order of magnitude higher than the macroscopic counterpart in the actuator.

As a measure of the effect of the nonlinear electro-mechanical coupling,
it is also demonstrated that the optimum configurations strongly depend on the maximum operational electric field chosen for the actuator. The softer is the matrix, the more observable is the phenomenon.

\vspace{5 mm}

{\bf Acknowledgements}. Support from the EU FP7 project ERC-AdG-340561-Insta\-bilities is gratefully acknowledged.

\section*{Appendix}

The strain energy proposed in \cite{spinelli_lopez_2015} to describe the electroelastic rank-one laminate response of the core relies on the following set of invariants:
$$
I_1=\bF_{\rm co}\scalp\bF_{\rm co},\ \ \ I_2=\bF_{\rm co}^{-T}\scalp\bF_{\rm co}^{-T},\ \ \ I_4=\bF_{\rm co} \bn^0_{\rm co}\scalp\bF_{\rm co} \bn^0_{\rm co},
$$
\beq
I_5=\bC_{\rm co}\bn^0_{\rm co}\scalp \bC_{\rm co}\bn^0_{\rm co},\ \ \ J_7=\bF_{\rm co}^{-T}\bE^0_{\rm co}\scalp\bF_{\rm co}^{-T}\bE^0_{\rm co},\ \ \ J_{10}=\bF_{\rm co}^{-T}\bE_{\rm co}^0\scalp \bF_{\rm co}^{-T}\bn^0_{\rm co},
\eeq
where $\bC_{\rm co}=\bF_{\rm co}^T \bF_{\rm co}$. Its expression is given by
\beq
W_{\rm co}=\frac{\bar \mu}{2} (I_1-3)-\frac{\bar \mu-\tilde \mu}{2}\left(I_4-\frac{1}{I_2-I_1 I_4+I_5}\right)-\frac{\bar \varepsilon}{2} J_7+ \frac{\bar \varepsilon-\tilde\varepsilon}{2} \frac{J_{10}^2}{I_2-I_1 I_4+I_5},
\eeq
where a superposed bar signifies arithmetic average, namely $\bar f=c^{\rm co}_a f_a+c^{\rm co}_b f_b$, while
$\tilde f^{-1}=c^{\rm co}_a/f_a+c^{\rm co}_b/f_b$.


\end{document}